# Entropy based Anomaly Detection System to Prevent DDoS Attacks in Cloud

A.S.Syed Navaz,      V.Sangeetha,      C.Prabhadevi

Department of Computer Applications,
Muthayammal College Of Arts & Science, Namakkal, India.

## ABSTRACT
Cloud Computing is a recent computing model; provides consistent access to wide area distributed resources. It revolutionized the IT world with its services provision infrastructure, less maintenance cost, data and service availability assurance, rapid accessibility and scalability. Grid and Cloud Computing Intrusion Detection System (GCCIDS) detects encrypted node communication and find the hidden attack trial which inspects and detects those attacks that network based and host based can't identify. It incorporates Knowledge and behavior analysis to identify specific intrusions. Signature based IDS monitor the packets in the network and identifies those threats by matching with database but It fails to detect those attacks that are not included in database. Signature based IDS will perform poor capturing in large volume of anomalies. Another problem is that Cloud Service Provider (CSP) hides the attack that is caused by intruder, due to distributed nature; cloud environment has high possibility for vulnerable resources. By impersonating legitimate users, the intruders can use a service's abundant resources maliciously. In Proposed System we combine few concepts which are available with new intrusion detection techniques. Here to merge Entropy based System with Anomaly detection System for providing multilevel Distributed Denial of Service (DDoS). This is done in two steps: First, Users are allowed to pass through router in network site in that it incorporates Detection Algorithm and detects for legitimate user. Second, again it pass through router placed in cloud site in that it incorporates confirmation Algorithm and checks for threshold value, if it's beyond the threshold value it considered as legitimate user, else it's an intruder found in environment. This System is represented and maintained by as third party. When attack happens in environment, it sends notification message for client and advisory report to Cloud Service Provider (CSP).

## 1. INTRODUCTION
Cloud Computing is being changed and altered to a new model consisting of services that are commoditized and delivered in a style similar to conventional utilities such as water, gas, electricity, and telephony service. In such a model, customers access services based on their requirements without gaze at to where the services are hosted or how they are delivered. Cloud computing denotes the infrastructure as a "Cloud" from which businesses and customers are competent and capable to access applications from anywhere in the world using on demand techniques.

The Cloud Computing Service Model is based on three primary tenants – Infrastructure as a Service (IaaS), Platform as a Service (PaaS) and Software as a Service (SaaS). All IT functions such as applications, networking, security, storage and software work in tandem to provide users with a service based on the client-server model

*1.1 Infrastructure as a Service* : The capability provided to the consumer is to provision processing, storage, networks, and other fundamental computing resources where the consumer is able to deploy and run arbitrary software, which can include operating systems and applications. The consumer does not manage or control the underlying cloud infrastructure but has control over operating systems, storage, deployed applications, and possibly limited control of select networking components (e.g., host firewalls).

*1.2 Platform as a Service* : The capability provided to the consumer is to deploy onto the cloud infrastructure consumer-created or acquired applications created using programming languages and tools supported by the provider. The consumer does not manage or control the underlying cloud infrastructure including network, servers, operating systems, or storage, but has control over the deployed applications and possibly application hosting environment configurations.

*1.3 Software as a Service* :  The capability provided to the consumer is to use the provider's applications running on a cloud infrastructure. The applications are accessible from various client devices through a thin client interface such as a web browser (e.g., web-based email). The consumer does not manage or control the underlying cloud infrastructure including network, servers, operating systems, storage, or even individual application capabilities, with the possible exception of limited user-specific application configuration settings.

*1.4 Deployment Models* : there are four deployment models for cloud services, with derivative variations that address specific requirements:

*1.4.1 Public Cloud* : The cloud infrastructure is made available to the general public or a large industry group and is owned by an organization selling cloud services.

*1.4.2 Private Cloud* : The cloud infrastructure is operated solely for a single organization. It may be managed by the organization or a third party, and may exist on-premises or off-premises.

*1.4.3 Community Cloud* : The cloud infrastructure is shared by several organizations and supports a specific community that has shared concerns (e.g., mission, security requirements, policy, or compliance considerations). It may be managed by the organizations or a third party and may exist on-premises or off-premises.

*1.4.4 Hybrid Cloud :* The cloud infrastructure is a composition of two or more clouds (private, community, or public) that remain unique entities but are bound together by standardized or proprietary technology that enables data and application portability (e.g., cloud bursting for load-balancing between clouds).

## 2. ISSUES
In general, security issues have been categorized into broad categories including sensitive data access, data segregation, privacy, bug exploitation, recovery, accountability, malicious insiders, management console security, account control, and multi-tenancy issues. Majority of the threats in the existing system arises from Service Oriented Architecture (SOA) ie combination of SOA and cloud computing which may expose security threats, and make controlling access to information potentially difficult. Low level of understanding can also generate threats.

Since Cloud computing supports distributed service oriented paradigm, multi-domain and multi-users administrative infrastructure, it is more prone to security threats and vulnerabilities. Currently the biggest hurdle in cloud adoption by most of the corporate organizations is its security. Due to its distributed nature, cloud environment has high intrusion prospects and suspect of security infringements. Large business organizations place there data into Cloud and get worry-free as a Cloud service provider (CSP), stores & maintains data, application or infrastructure of cloud user. The control over data and application poses the challenges of security like data integrity, confidentiality and availability.

DDoS: it is an attack where multiple compromised systems infected with a Trojans are used to target a single system causing a Denial of Service (DoS) attack. Victims of a DDoS attack consist of both the end targeted system and all systems maliciously used and controlled by the hacker in the distributed attack with a high impact on the service provider than the clients. These hazardous infections seriously affect the company reputation, client trust and interest.

## 3. Related Work and Existing System
*3 .1 Intrusion Detection for grid and Cloud computing  :* Each node identifies local events that could represent security violations and alerts the other nodes. Each individual IDS cooperatively participates in intrusion detection. The sharing of information between the IDS service and the other elements participating in the architecture: the node, service, event auditor, and storage service. Kleber, schulter et al. have proposed that IDS service increases a cloud's security level by applying two methods of intrusion detection. The behavior-based method dictates how to compare recent user actions to the usual behavior. The knowledge-based method detects known trails left by attacks or certain sequences of actions from a user who might represent an attack. The audited data is sent to the IDS service core, which analyzes the behavior using artificial intelligence to detect deviations. The analyzer uses a profile history database to determine the distance between a typical user behavior and the suspect behavior and communicates this to the IDS service. The rules analyzer receives audit packages and determines whether a rule in the database is being broken .It returns the result to the IDS service core. With these responses, the IDS calculate the probability that the action represents an attack and alerts the other nodes if the probability is sufficiently high.

## 4. Distributed Cloud Intrusion Detection Model
NIDS and HIDS are not suitable for security environment of cloud. Cloud as middleware layer, which having an audit system that design to cover an attacks that HIDS and NIDS cant cover. Irfan Gul,et al. have suggested So by means of using this model it can be able to bring the IDS as middle ware and any information from cloud user to CSP will reached through by means of it. This middleware is said to be as third party and it was fully maintained by service provider.

*4.1 Intrusion detection in the cloud :* Intrusion detection system plays an important role in the security and perseverance of active defense system against intruder. IDS implementation in cloud computing requires an efficient, scalable and virtualization-based approach. Sebastian Roschke et al.have proposed that cloud computing, user data and application is hosted on cloud service provider's remote servers and cloud user has a limited control over its data and resources. In such case, the administration of IDS in cloud becomes the responsibility of cloud provider. It provides integration solution for central IDS management. Deployment of IDS sensors on separate cloud layers like application layer, system layer and platform layer. Alerts generated are sent to "Event Gatherer" program. Event gatherer receives and convert alert messages in IDMEF standard and stores in event data base repository with the help of Sender, Receiver and Handler plug-ins.

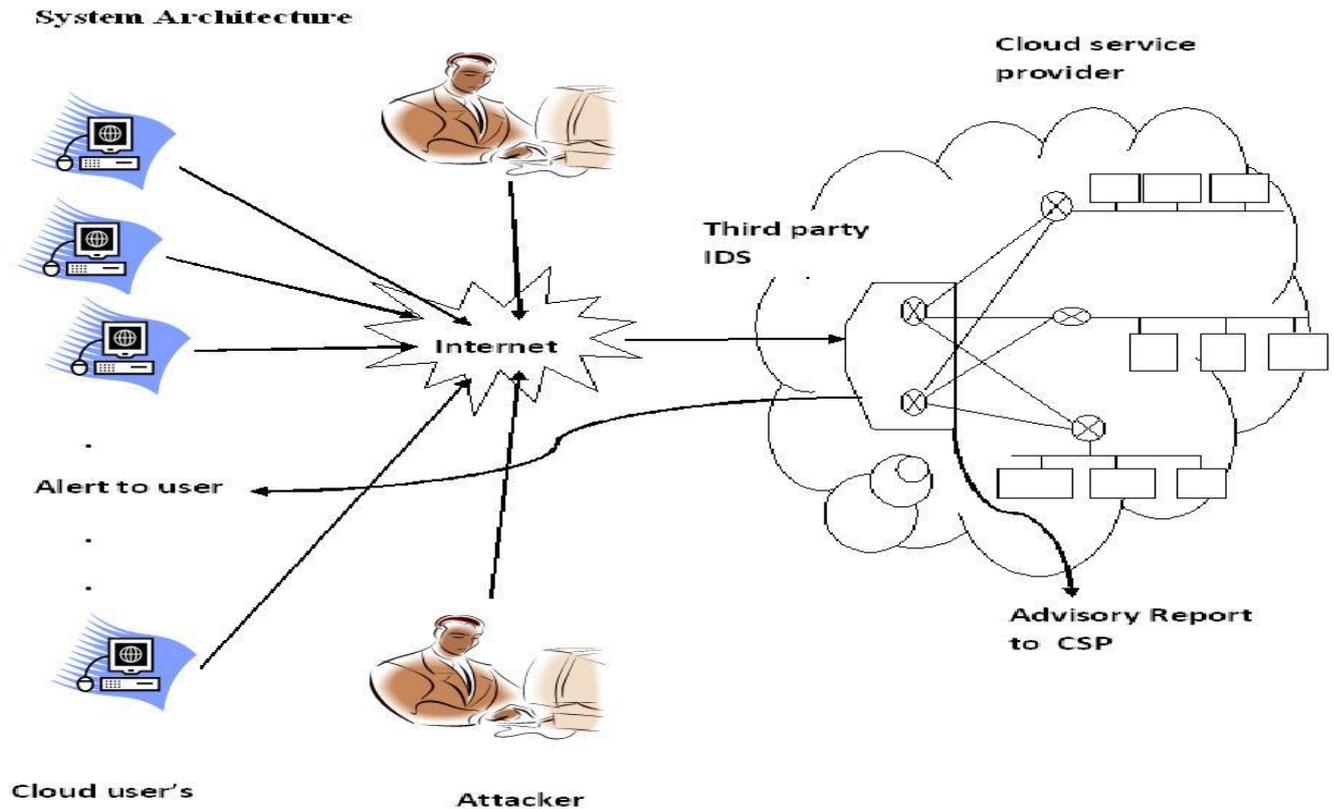

**Fig 1: System Architecture of Cloud service provider**

## 5. Cloud QoS, High Availability & Service Security Issues with Solutions

Distributed Denial of Service (DDoS) poses as a potential intimidation and danger to this key technology of the expectations and future. Muhammad Zakarya et al. have suggested a new Cloud Environment and Architecture and an Entropy based Anomaly Detection System (ADS) approach to mitigate the DDoS attack which further improves network performance in terms of computation time, Quality of Service (QoS) and High Availability (HA) under Cloud Computing environment. Entropy uses two algorithm to mitigate the intruders namely detection algorithm and confirmation algorithm.

## 6. Online Detection of Network Traffic Anomalies Using Degree Distributions

One problem is that the amount of traffic data does not allow real-time analysis of details. Another problem is that some generic detection metrics possess lower capabilities on diagnosing anomalies. To overcome these problems, Wuzuo WANG et al. have suggested an explicit algorithm to perform on-line traffic analysis is modeled. In this scheme, first make use of degree distributions to effectively profile traffic features, and then use the entropy to determine and report changes of degree distributions, which changes of entropy values can accurately differentiate a massive network event, normal or anomalous by adaptive threshold.

## 7. AN ANOMALY DETECTION SYSTEM FOR DDOS ATTACK IN GRID COMPUTING

Grid computing is rapidly emerging as a dominant field of wide area distributed computing. Grid computing is a collection of heterogeneous computers and resources across multiple organizations and delivers computing and resources as services to its users. The heterogeneity and scalability characteristics of Grid introduce potential security challenges. Distributed Denial of Service attack (DDoS) is one of the major threats to grid computing services. Sumit kar et al. have proposed the method to secure system for DDoS attack is based on the 3 steps: (i) Attack prevention, (ii) attack detection and recovery, and (iii) attack identification. This paper presents vulnerability of Grid computing in presence of DDoS attack. The proposed method is based upon attack detection and recovery, and uses an Entropy based anomaly detection system to detect DDoS attack. A grid topology model is used to describe how to implement the entropy based anomaly detection system in grid environment

*7.1 An Anomaly Detection Scheme for DDoS Attack in Grid Computing :* In this system they illustrated about deploying efficient intrusion detection system to Grid can significantly improve its security and it can detect denial of service attack before it affects the victim. But due to the special characteristics and requirement of Grids, the existing traditional intrusion detection system cannot work properly in that environment. Sumit kar focus of this thesis is to investigate and design an anomaly detection system which can detect DoS and DDoS attack with high attack detection and low false alarm rate to achieve high performance. Have to extensively surveyed the current literatures in this area; the main stress is put on feature selection for the Grid based anomaly detection system. An entropy based anomaly detection system has been proposed; also have discussed the advantage of taking entropy as the metric.

## 8. Integrating a Network IDS into an Open Source Cloud Computing Environment

The analyzed data source, IDS can be classified in network and host based. Network based IDS analyze traffic flowing through a network segment, by capturing packets in real time, and analyzing and checking them against some "classification" criteria. Claudio Mazzariello et al. have suggested that IDS can be further characterized with respect to the type of detection mechanism implemented. Namely, IDS can explicitly model attacks, anomalies and unwanted behavior, thus implementing the misuse-based detection paradigm, or conversely model normal and expected events, consequently detecting as anomalous what doesn't conform to such "normality" model.

*8.1 Mining Anomalies Using Traffic Feature Distributions :* Treating anomalies as events that disturb the distribution of traffic features differs from previous methods, Anukool Lakhina et al. have largely focused on traffic volume as a principal metric. In comparison, feature-based analysis has two key benefits. First, it enables detection of anomalies that are difficult to isolate in traffic volume. Some anomalies such as scans or small DOS attacks may have a minor effect on the traffic volume of a backbone link, and are perhaps better detected by systematically mining for distributional changes instead of volume changes. Second, unusual distributions reveal valuable information about the structure of anomalies information which is not present in traffic volume measures. The distributional structure of an anomaly can aid in automatic classification of anomalies into meaningful categories. This is a significant advance over heuristic rule-based categorizations, as it can accommodate new, unknown anomalies and at the same time expose their unusual features

8.2 ANOMALY : Maintaining security in Cloud Environment is a real challenge and most tough part in security management of large high speed networks like grid is the detection of suspicious anomalies in network traffic patterns due to DoS and DDoS attacks. To secure Cloud from DoS attack its must be detected before it affects the cloud user with high detection rate and low false rate, So that attack traffic will be discarded, without affecting legitimate traffic. In case of DDoS attack, the attack packets comes from ten or thousand of sources and DoS defense system that is based upon monitoring the volume of packets coming from a single address or single network an so it was get failed because the attack comes from various sources. Intrusion detection systems are widely used for DDoS attack detection. An intrusion detection system (IDS) inspects all inbound and outbound network and system activity and identifies possible security threats in a network or in a system. IDSs can be classified, based on their functionality, as misuse detectors and anomaly detectors. Anomaly detection has an advantage over signature-based is that a new attack can be detected if it falls out of the normal traffic patterns. Due to the special characteristics and requirement of computational grids, detecting such difference in traffic patterns imposed some new unique challenges that did not exist in traditional intrusion detection system. Anomaly Based detection is Useful because it can detects all kind of traffic behavior that is new or unusual, the anomaly-based method is brilliant for providing early warning for potential intrusions. These warnings can cover soundless attempts, backdoor activities, and certain natural failures in the network. Different techniques and challenges involved in anomaly detection system can be established. Many articles suggested that use of traffic volume [flow, packet, byte count] as the metric for anomaly detection system. In this system network is being monitored for potential security activites, an IDS was implemented by combining both Anomaly and Entropy intrusion detection system.The use of entropy for analyze the changes in traffic distribution which has two advantage i) Using entropy for anomaly detection which will increases the detection capability when compared with volume based methods. ii) Entropy method will provide additional information to categorize among dissimilar types anomaly (worms, DDoS attack scanning).

Considers two classes of distribution i) flow header features (IP address, ports, and flow sizes) ii) behavioral features (the number of distinct destination / source address that a host communicates with) .The attack discussed above can be better detected by analyzing distribution of traffic features. A traffic feature is a field in the header of the packet. The anomaly detection system discussed in this paper is based on by analyzing the change in entropy of above two traffic distributions.

### 8.3 ENTROPY BASED APPROACH :

Entropy or Shannon-Wiener index is an important concept of information theory, which is a measure of the uncertainty or randomness associated with a random variable or in this case data coming over the network. If it was more random it contains more entropy. The value of sample entropy lies in range [0, logn]. The rate of entropy is lesser when the class distribution is pure i.e.it belongs to one class. The rate of entropy is larger when the class distribution is impure i.e. class distribution belongs to many class. Hence comparing the rate of entropy of some sample of packet header fields to that of another sample of packet header fields provides a mechanism for detecting changes in the randomness.

The entropy shows its minimum value 0 when all the items (IP address or port) are same and its maximum value logn when all the items are different. Use change of entropy of traffic distributions (IP address, port) for DDoS detection. If you are interested in measuring the entropy of packets over unique source or destination address then maximum value of n is $2^{32}$ for ipv4 address. If you want to calculate entropy over various applications port then n is the maximum number of ports.

The entropy H (X) of a random variable X with possible values $\{x_1, x_2…, x_n\}$ and distribution of probabilities P = $\{p_1, p_2, . . . , p_n\}$ with n elements, where $0 <= p_i <= 1$ and

Here p (xi) where xi belongs to X is the probability that X takes the value xi. Suppose it randomly observe X for a fixed time window w, then p (xi) = mi/m, where mi is the frequency or number of times it observe X taking the value xi

**When to calculate probability of any source (destination) address then,**

$$P(x) = \frac{\text{Number of pkts with x as src (dst) address}}{\text{Total number of packets}}$$

$m_i$ = number of packets with xi as source (Destination) address ,m = total number of packets

Here total number of packets is the number of packets seen for a time window *T*.

**Similarly it can calculate probability for each source (destination) port as**

$$P(x) = \frac{\text{Number of pkts with X as src (dst) port}}{\text{Total number of pkts}}$$

**When to calculate with flow size**

$$P(x) = \frac{\text{Number of flows with flow size}}{\text{Total number of flows}}$$

Normalized entropy calculates the over all probability distribution in the captured flow
for the time window *T*.

**Normalized entropy = (H / log $n_0$)**

If NE<th1, th1 is threshold value1, Mark flow as suspected and raise an alert
Here $n_o$ is the number of distinct xi values in the given time window

In a DDoS attack from the captured traffic in time window *T*, the attack flow dominates the whole traffic, as a result the normalized entropy of the traffic decreased in a detectable manner. But it is also possible in a case of enormous genuine network accessing. To confirm the attack have to again calculate the entropy rate. Here flow is packages which share the same destination address/port. In this mechanism have taken one assumption that the attacker uses same function to generate attack packets at

*8.4 Projected Entropy:* According to for a stochastic processes the entropy rate H (x) of two random processes are same

$$H(\chi) = \lim_{n \to \infty} \frac{1}{n} H(x_1, x_2 \ldots x_n)$$

If H(x) <=th2, th2 is the threshold value2,Mark the flow as attacked, raise a final alert, discard the attack flow

## 9. Proposed System
In the proposed system was implement with the integrated technique such as Entropy based system and Anomaly detection system for providing Multilevel DDoS detection and prevention flow DDoS attack .To handle a large number of data packets flow in such an environment a multi-threaded IDS approach has been proposed in this paper. The multi-threaded IDS would be able to process large amount of data and could reduce the packet loss. After an efficient processing the proposed IDS would pass the monitored alerts to a third party monitoring service, who would in turn directly inform the cloud user about their system under attack. This System is represented as third party and that was maintained by CSP, and this will send message if any attack happens in the environment and also send a advisory report to Cloud service provider.

Entropy based System with Anomaly detection System for providing multilevel Distributed Denial of Service (DDoS). This is done in two steps: First, Users are allowed to pass through router in network site in that it incorporates Detection Algorithm and detects for legitimate user. Second, again it pass through router placed in cloud site in that it incorporates confirmation Algorithm and checks for threshold value, if it's less than the threshold value it considered as an intruder found in environment. This System is represented and maintained by as third party. When attack happens in environment, it sends notification message for client and advisory report to Cloud Service Provider (CSP).

Intrusion detection System administrator assigns a threshold value for packets. Collect traffic flow in time slot. Calculate Entropy H(x) for packets, by means of IP address, ports and flow size as input. From this, normalized entropy is found. Finally compare normalized entropy with that of assigned threshold value. If Normalized Entropy is smaller than threshold entropy then, the received packet is from illegal user else comparison is done against another threshold value. In this case larger value concludes that packet is received from legal user.setting a threshold value is not an easy task. Threshold value mainly depends on false positive rate.

This System works efficient in addition with degree distribution. Here for each value of in degree calculate entropy to diagnose the anomaly

## 11 . Author's Profile

**A.S.Syed Navaz** received BBA from Annamalai University, Chidambaram 2006, M.Sc Information Technology from KSR College of Technology, Anna University Coimbatore  2009, M.Phil in Computer Science from Prist University, Thanjavur 2010 and M.C.A from Periyar  University, Salem  2010 .Currently he is working as an Asst.Professor  in Department of BCA, Muthayammal College of Arts & Science, Namakkal. His area of interests are Computer  Networks and Mobile Communications.

**V.Sangeetha** received B.Sc Computer Science from Mahendra Arts & Science College, Periyar University 2003, M.Sc Computer Science from Mahendra Arts & Science College, Periyar University 2005 and M.Phil in Computer Science from Periyar  University, Salem  2008. Currently she is working as an Asst.Professor  in Department of BCA, Muthayammal College of Arts & Science, Namakkal. Her area of interests are Computer  Networks and Mobile Communications.

**C.Prabhadevi** received BCA from Sengunthar Arts & Science College, Periyar University 2005 , M.Sc Computer Science from Sengunthar Arts & Science College, Periyar University 2007, M.C.A from Periyar  University, Salem  2009 and M.Phil in Computer Science from Prist University, Thanjavur  2012. Currently she is working as an Asst.Professor  in Department of BCA, Muthayammal College of Arts & Science, Namakkal. Her area of interests are Digital Image Processing and Mobile Communications.